\documentclass[runningheads]{llncs}

\usepackage{graphicx}
\usepackage{makecell}
\usepackage{url}
\usepackage{xcolor}
\usepackage{color}
\usepackage{multirow}
\usepackage{amsmath}
\usepackage[normalem]{ulem}
\title{Brain Tumor Survival Prediction using Radiomics Features}
\author{Sobia Yousaf\inst{1} \and Syed Muhammad Anwar\inst{1} \and Harish RaviPrakash\inst{2} \and Ulas Bagci\inst{2}}
\date{\today}
\institute{Department of Software Engineering, UET Taxila, Taxila, Pakistan \and CRCV, University of Central Florida, Orlando FL, USA}

\begin{document}
\maketitle

\begin{abstract}
Surgery planning in patients diagnosed with brain tumor is dependent on their survival prognosis. A poor prognosis might demand for a more aggressive treatment and therapy plan, while a favorable prognosis might enable a less risky surgery plan. Thus, accurate survival prognosis is an important step in treatment planning. Recently, deep learning approaches have been used extensively for brain tumor segmentation followed by the use of deep features for prognosis. However, radiomics-based studies have shown more promise using engineered/hand-crafted features. In this paper, we propose a three-step approach for multi-class survival prognosis. In the first stage, we extract image slices corresponding to tumor regions from multiple magnetic resonance image modalities. We then extract radiomic features from these 2D slices. Finally, we train machine learning classifiers to perform the classification. We evaluate our proposed approach on the publicly available BraTS 2019 data and achieve an accuracy of \textbf{76.5\%} and precision of 74.3\% using the random forest classifier, which to the best of our knowledge are the highest reported results yet. Further, we identify the most important features that contribute in improving the prediction. 
\end{abstract}

\section{Introduction}
Gliomas are the most common type of brain tumor with over 78\% malignant tumors being gliomas~\cite{url4}. However, not all gliomas are malignant and can be broadly classified into two groups: 
high-grade glioma (HGG) and low-grade glioma (LGG). According to the World Health Organization guidelines four grades are defined for tumors~\cite{louis20162016}. Grade I and Grade II tumors are LGG, which are primarily benign and slow growing. Grades III and IV are HGG, which are malignant in nature with a high probability of recurrence. With Grade I tumors being mostly benign, patients tend to have a long term survival rate. Patients with HGG on the other hand, owing to the more aggressive nature of these tumors, have a much lower survival time, sometimes not exceeding a year. 
An early diagnosis of glioma would help the radiologist in assessing the patient's condition and plan a treatment accordingly. Magnetic resonance images (MRI) provide high contrast for soft tissue and hence represent the heterogeneity of the tumor core, providing a detailed information about the tumor. Different modalities commonly used in radiology include T1-weighted, contrast enhanced T1-weighted, fluid-attenuated inversion recovery (FLAIR), and T2-weighted MRIs.
A detailed profile of enhancing tumor region can be described using contrast enhanced T1-weighted MRI as compared to T2-weighted MRI \cite{villanueva2017current}. The hyper-intense regions in FLAIR images tend to correspond to regions of edema thus suggesting the need for the use of multi modal images~\cite{ho2012cerebral}.  
A quantitative assessment of the brain tumor provides important information about the tumor structure and hence is considered as a vital part for diagnosis \cite{sun2018tumor}. Automatic tumor segmentation of pre-operative multi modal MR Images in this perspective is attractive because it provides the quantitative measurement of the tumor parameters such as shape and volume. This process is also considered as a pre-requisite for survival prediction, because significant features can only be computed from the tumor region. So, this quantitative assessment has a significant importance in the diagnosis process and research. 
Due to imaging artifacts, ambiguous boundaries, irregular shape and appearance of tumor and its sub-regions, development of automatic tumor segmentation algorithms become challenging. 

Over the past few years, 
several deep learning (DL) based approaches have been introduced especially for medical image analysis\cite{anwar2018brain,anwar2019survey,raviprakash2020deep,mehreen2019hybrid}. 
DL models outperform 
traditional machine learning approaches on numerous applications in computer vision as well as medical image analysis, especially when sufficiently large number of training samples are available \cite{anwar2018medical}. 
Automatic segmentation alone is not sufficient for diagnosis and treatment, hence survival analysis is also necessary to help determine the treatment and therapy plans. Towards this, traditional machine learning based approaches have shown 
promising results using hand-crafted features~\cite{sun2018tumor}. These features, having been extracted from radiology images, are commonly referred to as radiomic features and help in the characterization of the tumor. Since 2017, the challenge of survival prognosis of glioma patients on BraTS benchmark data has been included. 
In the task of survival prediction, patients diagnosed with HGG are categorized into short-, mid- and long-term survival groups. The interval of these classes can be decided on the basis of number of days, months, or resection status. While DL algorithms have shown excellent performance on tumor segmentation tasks, in the survival prediction task, they have shown unstable performance~\cite{suter2018deep}. Radiomics are likely to be dominant for precision medicine because of its capability to exploit detailed information of gliomas phenotype \cite{weninger2018segmentation}. Inspired from these achievements of radiomics features in this challenging task of survival prediction on several modalities, herein we propose to utilize radiomic features for prediction.

\section*{Our Contributions}
In this paper we present a machine learning based approach, utilizing radiomic features, for survival prediction on BraTS 2019 data. We extract radiomic features from the tumor regions utilizing the provided ground-truth segmentation masks and train machine learning classifiers to predict the survival class. Our main contributions are
\begin{itemize}
    \item We identify discriminating features that contributed the most in improving the accuracy and found that Haralick features are more significant for survival prediction task. 
    \item We explore multiple classifiers commonly used in this domain, and found random forest to be the best performing model with state-of-the-art performance when used with the selected radiomics features.
   
\end{itemize}

\section{Proposed Methodology}
\begin{figure*}[!ht]
\centering 
\includegraphics[width = 120mm] {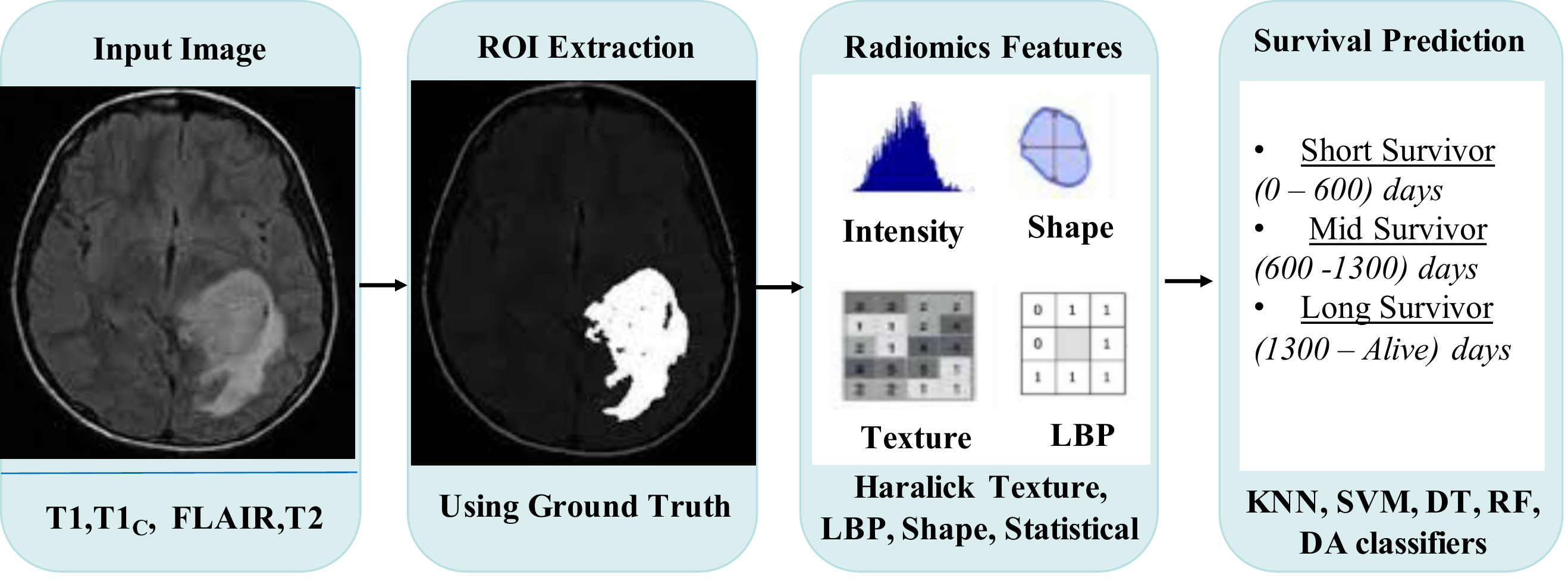}
\caption{\footnotesize{The proposed radiomics features based survival prediction pipeline using BraTS 2019 data.}}
\label{fig:res1}
   \end{figure*}

Our proposed approach towards survival prediction is shown in Fig. \ref{fig:res1}. It consists of the three main steps. 1) Region of interest (ROI) extraction 2) Radiomic features computation, and 3) Survival prediction 
The details of these steps are presented in the following sections.

\vspace{-10mm} 
\subsection{Image pre-processing and ROI extraction}
The data were acquired from multiple institutions using different scanners. There could exist different levels of noise in scanners leading to intensity variations that can strongly influence the extracted radiomic features~\cite{bakas2017advancing}. Hence, bias field correction and normalization steps were applied on the input data to standardize the intensity values. More precisely, the intensity value of each image slice is subtracted from its mean and is divided by the image’s intensity standard deviation. In order to extract the ROI, we applied the ground truth of respective patient on all input modalities. As a result we get the complete tumor region from all scans of the patient.
\subsection{Radiomics Features}
Radiomics are the specific kind of features that are primarily computed from radiology images to describe phenotypes of the tumor region. These features can be further used to predict the tumor and can improve the survival prediction. 
Herein, radiomic features extracted from input modalities are classified into three groups
- first order statistics, shape features, and texture features. The details of these features are presented in the following text. 
\vspace{-5mm}
\subsubsection*{First-order statistics}
These features represent statistical properties such as the average intensity value, median, variance, standard deviation, kurtosis, skewness, entropy, and energy. These features were computed using the intensity values in MR images, such that the gray-level intensity of the tumor region is described accurately. 
In particular, a total of 10 first order features were extracted from each slice of the four modalities used.

\vspace{-5mm}
\subsubsection{Shape Features}
Shape features 
include perimeter, area, convex area, convex perimeter, concavity, diameter, major and minor axis length, circulatory, elongation, and sphericity. Further, we described the tumor shape by using the Fourier descriptor, 
where the entire shape is represented using minimal numeric values \cite{burger2013fourier}. 

\vspace{-5mm}
\subsubsection{Texture Features}
Texture features are considered to be strong in the radiomics field~\cite{yang2015evaluation}. We computed the Haralick texture features \cite{haralick1973textural}, and local binary patterns (LBP). In particular, Haralick features were computed from the gray level co-occurrence matrix (GLCM), which describes the spatial relationship among pixels. 
Whereas in LBP, a binary encoded representation is used to describe the relationship between pixels of interest with its neighbors~\cite{polepaka2019idss}. A total of 14 Haralick features (shown in Table \ref{tab:my-table}) and 55 LBP features were extracted from each slice.
In particular, G represents the number of gray levels, i and j are indices of the pixels of these gray levels, while $P(i,j)$ denotes the intensities of the pixel in the GLCM matrix. While $\mu$, $\sigma$, and $\sigma^2$ represent the mean, standard deviation, and variance respectively.

\begin{table}[!th]
\caption{Description of Haralick Texture Feature's used in this study for survival prediction.}
\centering
\scalebox{0.70}{

\begin{tabular}{|l|l|}
\hline
\multirow{2}{*}{\textbf{Feature Name}} & \multicolumn{1}{c|}{\multirow{2}{*}{\textbf{Equation}}} \\
 & \multicolumn{1}{c|}{} \\ \hline
Probabilities P(x), P(y) & \parbox[c][30pt]{0.8\textwidth}{\begin{equation*}
P_x(i) = \sum_{i=0}^{G-1} P(i,j),  P_y(j) = \sum_{j=0}^{G-1} P(i,j)
\end{equation*}} \\ \hline
Variance & \parbox[c][30pt]{0.8\textwidth}{\begin{equation*}
Var =\sum_{i=0}^{G-1}\sum_{j=0}^{G-1} (i-\mu) P(i*j)
\end{equation*}} \\ \hline
Standard Deviation & \parbox[c][30pt]{0.8\textwidth}{\begin{equation*}\sigma_x^2(i) = \sum_{i=0}^{G-1} (P_x(i) - \mu_x(i))^2, \sigma_y^2(j) = \sum_{j=0}^{G-1} (P_y(j) - \mu_y(j))^2 \end{equation*}} \\ \hline
Homogeneity & \parbox[c][30pt]{0.8\textwidth}{\begin{equation*}
H = \sum_{i=0}^{G-1}\sum_{j=0}^{G-1} [P_x(i,j)]^2\end{equation*}} \\ \hline
Contrast & \parbox[c][30pt]{0.8\textwidth}{\begin{equation*}
C = \sum_{i=0}^{G-1} n^2 {\sum_{i=1}^{G}\sum_{j=1}^{G}} P(i,j), |i-j|=n \end{equation*}} \\ \hline
Correlation  & \parbox[c][30pt]{0.8\textwidth}{\begin{equation*}
Corr =\sum_{i=0}^{G-1}\sum_{j=0}^{G-1} (i*j)*P(i,j) - (\mu_x * \mu_y)/(\sigma_x * \sigma_y)
\end{equation*}} \\ \hline
Inverse Difference Moment & \parbox[c][30pt]{0.8\textwidth}{\begin{equation*}
IDM=\sum_{i=0}^{G-1}\sum_{j=0}^{G-1} P(i,j)/1+(i-j)^2
\end{equation*}} \\ \hline
Entropy & \parbox[c][30pt]{0.8\textwidth}{\begin{equation*}
Ent =\sum_{i=0}^{G-1}\sum_{j=0}^{G-1} P(i*j) * log(P(i,j))
\end{equation*}} \\ \hline
Average Sum & \parbox[c][30pt]{0.8\textwidth}{\begin{equation*}
S_A =\sum_{i=0}^{2G-2} i X P_x+y(i)
\end{equation*}} \\ \hline
Entropy Difference & \parbox[c][30pt]{0.8\textwidth}{\begin{equation*}
D_{Ent} = - \sum_{i=0}^{2G-2} P_x+y(i) * log(P_x+y(i))
\end{equation*}} \\ \hline
Entropy Sum & \parbox[c][30pt]{0.8\textwidth}{\begin{equation*}
S_{Ent} = - \sum_{i=0}^{G-} P_x+y(i) * log(P_x+y(i))
\end{equation*}} \\ \hline
Intertia &  \parbox[c][30pt]{0.8\textwidth}{\begin{equation*}
Inr =\sum_{i=0}^{G-1}\sum_{j=0}^{G-1} (i-j)^2 * P(i*j)
\end{equation*}}\\ \hline
\end{tabular}
}
\label{tab:my-table}
\end{table}

\subsection{Classification Models for Survival Prediction}
The survival prediction task is an important but challenging task for BraTS data. One of the reasons could be that only age and MR images are provided, hence this prediction mainly relies on tumor identification within the MRI. To this end, we have used radiomics features for describing tumors within the region of interest. In particular, we computed 90 features (statistical, shape- and texture-based) per subject per slice. These features are computed for complete tumor and fed to five different classifiers including discriminant analysis (DA), decision tree (DT), K-nearest neighbor (k-NN), support vector machine (SVM), and random forest (RF). 

k-NN is a simple machine learning algorithm that takes the all data available against defined classes and categorizes the incoming test samples on the basis of distance function or similarity measures. In our experiment, we have used different values of $k$ to evaluate the performance. 
DA is a statistical approach to find similar patterns or feature combinations to separate two or more data samples. The resultant combination of patterns can be used as a classifier to allocate samples to classes. 
SVM is a supervised machine learning model which maximizes the hyper-plane margin between different classes. The classifier maps input space into a high-dimension linearly separable feature space. Because of the nonlinear problem space we used the radial basis kernel. 
A DT starts dividing the data into smaller segments, meanwhile the tree is developed incrementally. The final tree contains two types of nodes i.e., decision and leaf nodes. Here every decision node has more than one branches (i.e. low, mid and long survivor) while the leaf node represents the final decision. In particular, we used 10 splits for the DT model. 
RF is one of the famous machine learning classifiers that is considered to be the best in response to over-fitting problems in large dimensional data. A RF model comprises of several trees that take random decisions on given training samples. For survival prediction, we used an RF with 30 bags and observed that accuracy increased with increasing the number of trees until it plateaued out. 
    
\section{Experimental Results}

\subsection{Dataset}
To evaluate the performance of our proposed method we used BraTS 2019 benchmark data provided by the Cancer Imaging Archive. The dataset comprises of independent training and validation sets. The training data contains $259$ subjects diagnosed with HGG and 76 subjects diagnosed with LGG along with ground truth annotations by experts. Moreover, the data comprises of MRI images from $19$ different institutions of four MRI modalities (T1-weighted, T2-weighted, T1-contrast enhanced and FLAIR). We selected the CBICA, BraTS 2013 and a single dataset from the TCIA archive resulting in 166 subjects with HGG. The images are pre-processed via skull-stripping, co-registration to a common anatomical template and re-sampling to an isotropic resolution of $1 \times 1 \times 1$ $mm^3$. 
The data also includes the survival information in terms of number of days for each patient along with their age.
In the BraTS 2019 data, the age range of the HGG cases is from 19 to 86 years and survival information ranges from 0 to 1767 days. It should be noted, that for some patients the survival information was missing, and we treated those as having low survival.
\subsection{Survival Prediction Performance}
We used the extracted radiomics features combined with clinical features to predict the survival class. All radiomics features were combined with patient age and hence a total of 30632 feature values were obtained from 166 HGG subjects to train five different conventional machine learning classifiers. 
These included a total of 90 radiomics features extracted per slice per subject, while no feature reduction technique was used. We chose five ML classifiers to evaluate the performance of the extracted radiomics features.  We created the class labels by normalizing and dividing the number of survival days into three different regions i.e., short survivor (0 – 600), medium survivor (600 – 1300), long survivor (1300 – Alive). The measurement criteria followed in literature is to predict the correct number of cases that has survival less than 10 months, between 10 to 15 months and greater than 15 months. We further used precision, recall and accuracy as performance measures for each classifier. Initially, we used first order statistical features and shape-based features, but found that these features could not provide a significant performance in the prediction task. Hence, we incorporated Haralick texture features and Fourier shape descriptor, and observed a significant increase in performance when using conventional classifiers. We used a 10-fold cross-validation approach for classification purpose.

\begin{table}[!t]
\centering
\caption{Performance evaluation of different classifiers for survival prediction using radiomics features. The bold values shows the best results. }
\label{tab:tab2}
\scalebox{0.8}{
\begin{tabular}{|c|c|c|c|}
\hline
\multirow{2}{*}{\textbf{Classifier}} & \multicolumn{3}{c|}{\textbf{Evaluation Metrics}}         \\ \cline{2-4} 
                                     & \textbf{Accuracy} & \textbf{Precision} & \textbf{Recall} \\ \hline
k-NN &0.388  &0.379  &0.365 \\ \hline
DA &0.471  &0.409  &0.399  \\ \hline
DT &0.678  & 0.640 &0.659   \\ \hline
SVM &0.526 & 0.509  & 0.519 \\ \hline
\textbf{RF} &\textbf{0.765}  & \textbf{0.743} & \textbf{0.736} \\ \hline
 FCNN \cite{wang20193d}  &0.515 & -  & - \\ \hline
\end{tabular}
}
\end{table}

Table \ref{tab:tab2} shows the performance of machine learning models using accuracy, precision, and recall parameters. 
A fully connected neural network, with two hidden layers, was used for survival prediction on Brats 2019 training data \cite{wang20193d}. For 101 patients, using radiomics features, an accuracy of $0.515\%$ was achieved. \textcolor{black}{We observed that k-NN shows poor performance with k=3, while RF gives the highest performance with 30 number of bags. The value of $k$ is an important parameter to choose for the k-NN classifier and impacts the overall classification results. Since k-NN performance was not at par when using radiomics features at $k=3$, we experimented with increasing the value of k, but did not observe a significant improvement in the performance. Our results indicate that random forest was able to learn the data representation from the radiomics features for overall survival prediction. In RF, each tree (total 30 trees) was diverse because it was grown and unpruned fully that's why the feature space was divided into smaller regions. Hence RF learned using the random samples, where a random feature set was selected at every node giving diversity to the model.} 

\begin{figure*}[!ht]
\centering 
\includegraphics[width = 100mm]{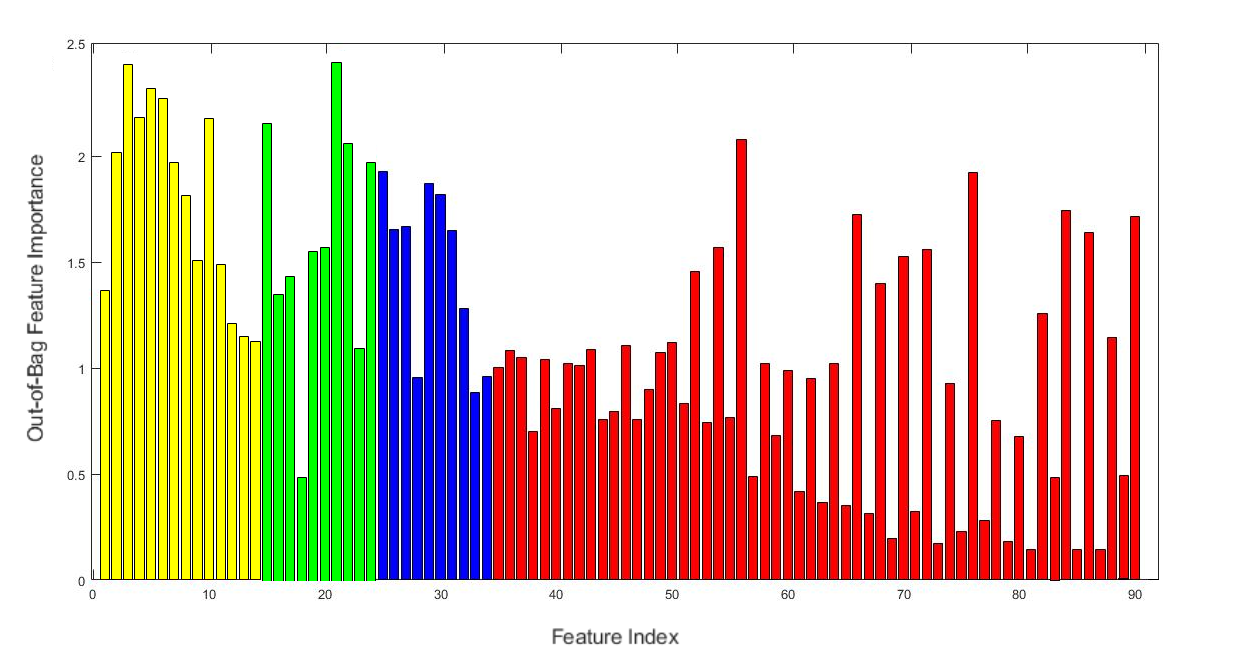}
\caption{\footnotesize{A representation of out-of-bag feature importance for all radiomics features used in this study with different colors, Haralick features(yellow), first order statistics features(green), shape features (blue) and LBP (red).}}
\label{fig:res2}
   \end{figure*}

\textcolor{black}{Since in BraTS 2019 benchmark data, the input modalities have intensity variations and tumor appearance is also heterogeneous, features computed from these modalities are also diverse in nature. We further quantify the importance of all features (statistical, shape-based, and Haralick) as shown in Figure \ref{fig:res2}. It was observed that Haralick features (represented on feature index 1 to 14) had an out-of-bag feature importance value ranging between 1 - 2.5. This was on average higher than all other set of features used and shows the significance of these features in the classification task. 
This analysis was performed in MATLAB using statistical and machine learning toolbox.
A confusion matrix for the best performing classifier (RF) is shown in Table \ref{tab:tab3}, where the values represent percentages.}

\begin{table}[!t]
\centering
\caption{Confusion matrix for random forest classifier (the values represent percentages).}
\label{tab:tab3}
\scalebox{0.8}{
\begin{tabular}{|l|l|l|l|}
\hline
\multirow{2}{*}{\textbf{Predicted Class}} & \multicolumn{3}{c|}{\textbf{Actual Class}} \\ \cline{2-4} 
 & Low Survival & Mid Survival & Long Survival \\ \hline
Low Survival & 76.52  & 9.04 & 14.44 \\ \hline
Mid Survival & 24.77 & 75.23  & 0\\ \hline
Long Survival & 19.00 & 6.38 & 74.62 \\ \hline
\end{tabular}
}
\end{table}

\vspace{-5mm}
\section{Discussion and Conclusion}

\vspace{-4mm}

\textcolor{black}{In this paper, we presented an automatic framework for the prediction of survival in patients diagnosed with glioma using multi modal MRI scans and clinical features. First, ROI radiomics features were extracted, which were then combined with clinical features to predict overall survival.
For survival prediction, we extracted shape features, first order statistics, and texture features from the segmented tumor region and then used classification models with 10-fold cross validation for prognosis.} 
In particular, the experimental data were acquired in multi-center setting and hence a cross-validation approach was utilized to test the robustness of our proposed approach in the absence of an independent test cohort. In literature, survival prediction model has been applied on diverse data along with different class labels and resection based clinical feature. For brain tumor, the performance in survival prediction has been lower, for instance an accuracy of $70\%$ was achieved \cite{sun2019brain}. 
In particular, 3D features were extracted from the original images and filtered images. Further, feature selection was performed to reduce the 4000+ features down to 14. While, in our proposed approach we utilize slice-based features (90) and majority voting across slices to obtain a final classification.  
\textcolor{black}{Among five classifiers mentioned above, RF showed the best results using the computed radiomics features. The performance significantly varied among these classifiers, which shows the challenging nature of this prediction. With RF, an accuracy of 0.76, along with precision and recall of 0.74 and 0.73, respectively was achieved. We also predicted subject wise evaluation of RF model where majority voting among slices from each patient was used to assign one of the three classes to the patient. We achieved an accuracy of $0.75$ using the subject-wise approach.
In future, we intend to extend this work by incorporating more data from the TCIA archive as well as using 3D features extracted from atlas based models for survival prediction. We also intend to use Cox proportional hazards models to better handle data with no survival information provided (missing data).} 

\vspace{-5mm}

\bibliographystyle{splncs04}

\bibliography{ref}
\end{document}